\documentclass[12pt,preprint]{aastex}

\shortauthors{Knigge, C. et al.}
\shorttitle{The hidden white dwarf in DW~UMa}

\begin{document}

\title{Time-Resolved Ultraviolet Spectroscopy of the SW~Sex Star
DW~UMa: Confirmation of a Hidden White Dwarf and the UV
Counterpart to Phase 0.5 Absorption Events
\footnote{Based on observations with the NASA/ESA Hubble Space
Telescope, obtained at the Space Telescope Science Institute, which is
operated by the Association of Universities for Research in Astronomy,
Inc. under NASA contract No. NAS5-26555.}}

\author{Christian Knigge}
\affil{School of Physics \& Astronomy,
University of Southampton,
Southampton SO17 1BJ, UK}
\email{christian@astro.soton.ac.uk}

\author{Sofia Araujo-Betancor}
\affil{Space Telescope Science Institute, 
3700 San Martin Drive,
Baltimore, MD 21218, USA}
\email{araujo@stsci.edu}

\author{Boris T. G\"{a}nsicke}
\affil{Department of Physics, 
University of Warwick, 
Coventry CV4 7AL, UK}
\email{Boris.Gaensicke@warwick.ac.uk}
 
\author{Knox S. Long}
\affil{Space Telescope Science Institute, 
3700 San Martin Drive,
Baltimore, MD 21218, USA}
\email{long@stsci.edu}

\author{Paula Szkody}
\affil{Astronomy Department, University of Washington, Seattle, WA
98195, USA}
\email{szkody@astro.washington.edu}

\author{D. W. Hoard}
\affil{Spitzer Science Center, California Institute of Technology,
Mail Code 220-6,  1200 East California Boulevard, 
Pasadena, CA 91125, USA}
\email{hoard@ipac.caltech.edu}

\author{R. I. Hynes
\footnote{Hubble Fellow}}
\affil{Astronomy Department and McDonald Observatory, 
University of Texas at Austin, 1 University Station C1400, 
Austin, TX 78712, USA}
\email{rih@astro.as.utexas.edu}

\and

\author{V. S. Dhillon}
\affil{Department of Physics and Astronomy, University of Sheffield,
Sheffield S3 7RH, UK}
\email{vik.dhillon@sheffield.ac.uk}

\begin{abstract}

We present time-resolved, ultraviolet (UV) spectroscopy of the SW
Sex star DW~UMa in the high state. We confirm that, shortward of
1500~\AA, the high-state, UV continuum level is lower than the white
dwarf (WD)-dominated low-state level. We also do not see the WD 
contact phases in the high state eclipse light curves. These results
confirm our earlier finding that the WD in this system is hidden from 
view in the high state. Based on this, we caution that eclipse mapping of 
high-inclination SW Sex stars in the high state may yield incorrect or
misleading results. In the context of DW~UMa, we demonstrate explicitly that 
distance estimates obtained by recent eclipse mapping studies cannot
be reconciled with the WD-dominated low-state spectrum. We also show that the 
fluxes of the UV emission lines in the high state drop near orbital
phase 0.5. This is the first detection of a UV counterpart to the
class-defining phase 0.5 absorption seen in the optical emission lines
of SW~Sex stars.

\end{abstract}

\keywords{accretion, accretion disks ---
binaries: close --- novae, cataclysmic variables --- stars:
individual: DW~UMa}

\section{Introduction}
\label{introduction}

The SW Sex stars are a sub-class of nova-like cataclysmic variable
stars (CVs). As such, they are interacting binary systems containing a
Roche-lobe filling secondary star that is losing mass to an accretion
disk around a white dwarf (WD) primary. The SW Sex classification was
first suggested by Thorstensen et al. (1991), building on earlier work
by Downes et al. (1986), Honeycutt, Schlegel \& Kaitchuck (1986),
Shafter, Hessman \& Zhang (1988), and Szkody \& Pich\'{e} (1990). The
key common features of the four founding members of the class -- SW
Sex, DW UMa, V1315 Aql and PX And -- were transient, central absorption
features in the Balmer and He~{\sc i} lines near orbital phase
0.5. Also, the radial velocity curves derived from these lines
generally showed significant phase lags with respect to the orbital
motion of the WD. Other common spectroscopic properties within the class are 
single-peaked optical emission lines at orbital phases well
away from 0.5 (rather than the double-peaked lines expected from
edge-on accretion disks); weak eclipses of the Balmer and He I lines;
and He~{\sc ii} features that do appear to track the WD motion 
and are more deeply eclipsed than the low-ionization lines. 

The continuum eclipses of high-inclination SW Sex stars are also
unusual. They are considerably more V-shaped than those of "ordinary"
eclipsing, non-magnetic nova-like CVs (such as UX UMa). If the
eclipsed light is assumed to be produced by a geometrically thin, 
optically thick disk, 
this immediately implies that the outer disk contributes relatively
more light in SW Sex stars than in other CVs. Unsurprisingly, eclipse
mapping experiments have therefore invariably inferred very flat
surface brightness (and hence effective temperature) distributions
across the disks of SW Sex stars (e.g. Rutten, van Paradijs \&
Tinbergen 1992; Baptista, Steiner \& Horne 1996; Baptista et al. 2000;
Groot, Rutten \& van Paradijs 2001). This contrasts starkly with the
theoretically expected temperature distribution, which drops as
$R^{-3/4}$~with radius.

Against this background came the first time-resolved, ultraviolet
(UV) observations of a bona-fide SW Sex star --- DW UMa -- with the
{\em Hubble Space Telescope} (Knigge et al. 2000; Araujo-Betancor et
al. 2003). These fortuitously
occurred during a low state of the system, during which the optical
light was suppressed by approximately 3 mags relative to the normal high
state. Such low states are thought to be caused by a temporary
reduction or cessation of the mass supply from the donor star. The
UV observations were broadly consistent with this scenario, inasmuch
as the low state UV spectrum appeared to be dominated by the hot
($T_{eff} \simeq 50,000$~K) WD primary. However, a surprise was that
the far-UV flux level was higher in the low-state HST data than
in high-state UV observations obtained with 
the {\em International Ultraviolet Explorer} (IUE). This implies 
that the WD in DW~UMa is hidden from view in the high state. We
therefore suggested that the disk in this high-inclination system ($i 
\simeq 82^{o}$; Araujo-Betancor et al. 2003) is self-occulting in
the high state, i.e. that the disk rim obscures our view of the WD and
the central disk regions. 

If the WD and inner accretion flow in high-inclination SW Sex stars
are permanently occulted, then this needs to be accounted for in any
realistic model for these systems. After all, 
the majority of confirmed SW~Sex stars are eclipsing
\footnote{See DWH's {\em Big List of SW Sex Stars}~at 
{\url http://spider.ipac.caltech.edu/staff/hoard/biglist.html}.},
which indicates that geometric effects influence our ability to even
recognize members of the class. It is therefore important to confirm the 
result of Knigge et al. (2000), not least to rule out
cross-calibration errors between the high-state IUE data and the low
state-HST observations.

Here, we present a first look at time-resolved, HST observations of DW
UMa in the high state, which were obtained using exactly the same
observational set-up as the low-state data. The results confirm
conclusively that the WD in this system is hidden from view in the
high state and show, for the first time, that the UV lines take part
in the phase 0.5 absorption events that are the key optical
characteristic of the SW~Sex class.

\section{Observations and Analysis}
\label{observations}

Our high-state HST observations of DW~UMa took place on April 4, 2004
and covered just over two complete cycles of DW~UMa's 3.28~hr orbital
period. We used the acquisition image (obtained with the
STIS CCD and the F28X50LP filter) to estimate an optical (roughly
R-band) magnitude of 14.1. Thus DW~UMa was clearly in its normal high
state during these observations (see Stanishev et 
al. [2004] for a long-term light curve). Our UV spectroscopy was
carried out using the 52\arcsec~$\times$~0\arcsec .2 slit, the FUV-MAMA
detector, and the G140L grating. This combination covers 1150~\AA~--
1720~\AA~at a resolution of $\simeq 1$~\AA~(FWHM). TIME-TAG mode was
used throughout, allowing us to split the data into suitable
subexposures a posteriori. These subexposures were then calibrated
using the STIS pipeline, using the reference files
available in June 2004. For consistency in comparing low-state and
high-state data, we also recalibrated the low-state observations 
obtained in 1999 with the latest version of the pipeline and reference 
files. The absolute flux scale of our spectra should be accurate to
about 4\%.
  
Calibrated UV spectra from both states are shown and compared
in Figure~1. Both spectra were extracted from data taken
well-away from eclipse when the flux was relatively constant (both 
high and low state display considerable UV variability). The overall 
appearance of the high-state HST spectrum -- line-dominated, with a
flat continuum -- is very similar to that of the high-state IUE
spectrum shown by Knigge et al. (2000). The continuum flux levels are
also similar, and we find again that, shortward of 1500~\AA, the
low-state UV spectrum has a higher continuum flux level than the
high-state spectrum. In fact, at the shortest wavelength, the ratio of
low-state to high-state continuum flux exceeds a factor of four. Thus
even allowing for the possibility that our view of the lower half of
the WD might be blocked by an optically thick disk in the high state (and
that this disk might somehow avoid radiating in the UV), the
high-state UV flux is inconsistent with the assumption that our view
of the WD at the centre of the disk is unimpeded.

In Figure~2, we additionally show continuum light curves extracted
from our TIME-TAG data, focusing particularly on the eclipses. The
data were phase-folded using the ephemeris of Stanishev et 
al. (2004), and the vertical lines mark the WD contact phases measured
by Araujo-Betancor et al. (2003). There are no obvious, repeatable
steps in the light curves at or near the predicted WD ingress/egress
intervals. This is consistent with the low UV flux level and confirms
that the WD in DW~UMa is hidden in the high state.

The top panel of Figure~2 also shows the continuum-subtracted light
curve of the strongest line in the spectrum, C~{\sc iv}. This reveals
that (i) the line is much more weakly eclipsed than the adjacent
continuum, and (ii) in both orbital cycles, the line flux reaches a
minimum near phase 0.5 (the continuum light curve actually appears to
display a similar morphology in the second cycle). In order to allow a closer
examination of these effects, we present in Figure~3 the high-state UV
spectra binned in orbital phase. The key point to note from this is that the line
fluxes of all strong UV transitions are suppressed near orbital phase
0.5. This is the first  detection of a UV counterpart to the
class-defining phase 0.5 absorption seen in the low-ionization optical  
emission lines of SW Sex stars. The effect appears to be weakest in He~{\sc
ii}~1640~\AA, which is consistent with the absence or weakness of
0.5 absorption in the He~{\sc ii}~4686~\AA~lines of SW Sex stars. 

\section{Discussion and Conclusions}

Our UV observations establish conclusively that the WD in DW UMa is 
hidden from view in the normal high state. We still believe that the
simplest way to explain this finding is to postulate the existence of
a self-occulting accretion disk in this system. However, independent
of any specific theoretical framework, it 
is clear that the accretion flow cannot even approximately
be described as 2-dimensional, and that any viable model or
description of the system must account somehow for the observed
occultation. It is therefore more difficult to use the 
eclipses to gain insight into the nature of the high-state accretion
flow (but not impossible: see Hellier \& Mason [1989] for an example
in the context of low-mass X-ray binaries). By contrast, all 
existing high-state eclipse mapping studies of DW~UMa (and other SW~Sex
stars) have assumed that the disks in these systems are fully visible.
\footnote{In their eclipse mapping study, Stanishev et
al. (2004) comment on the occultation problem and suggest that the
optically thick base of an accretion disk wind may hide the WD in
DW~UMa. But this scenario is just a specific incarnation
of a self-occulting disk atmosphere with unknown geometry, and thus 
still violates the assumption of full disk visibility. It is also worth
recalling here the work of Rutten (1998), who shows that the
application of a standard (flat-disk) eclipse mapping code to a flared
disk can give reasonable results, {\em as long as the disk is not
self-occulting}. Once the inner disk is no longer visible, the
method produces artificially flat radial temperature distributions.} 
In our view, the immediate corollary of this is that results inferred
from these studies should be regarded with caution. 

We suspect that the short distance estimates for DW UMa suggested in
recent eclipse mapping studies ($d \simeq 300$~pc; B{\'{\i}}r{\' o}
2000; Stanishev et al. 2004) are a manifestation of this
problem. Such a short distance is inconsistent with the estimate 
d~=590~$\pm$~100~pc obtained from fitting the low-state spectrum with
a WD model atmosphere and also with the estimate d~=~930~$\pm$~160~pc 
obtained from Bailey's method (Araujo-Betancor et al. 2003). We have
considered whether the WD-dominated low-state spectrum can in some way
be reconciled with a distance of 300~pc. We thus show in Figure~1
three $\log{g} = 8$~DA 
WD models (calculated as in G\"{a}nsicke, Beuermann \& de Martino [1995])
for different effective temperatures ($T_{eff}$). The models have been
scaled to match the observed low-state spectrum at 1480~\AA, have not
been convolved with the instrumental line spread function and have
also not been rotationally broadened. Thus the width of Lyman~$\alpha$
in the models is a lower limit.

The $T_{eff} = 50,000$~K model was chosen to represent the
best-fitting WD model found by Araujo-Betancor et al. (2003). This
model fits the data fairly well away from metal lines and requires a
distance (d = 540~pc) in line with our previous estimate (this assumes
a fully visible WD and the best-bet WD radius found by Araujo-Betancor
et al.). By contrast, the eclipse-mapping distance of d = 300~pc
requires a WD model with $T_{eff} \simeq 28,000$~K. This model
produces much too wide a Lyman~$\alpha$~feature and can be ruled out
immediately. If only half the WD were visible in the low state
(e.g. because the lower half is blocked by a disk), d = 300~pc
corresponds to $T_{eff} \simeq 37,000$~K. This model still
overpredicts the width of Lyman~$\alpha$~significantly and can also be
ruled out.\footnote{We actually do not consider this to be a
physically realistic model: if an optically thick disk were
present in the low state, it would be strongly irradiated and
therefore hot. It should then emit copious amounts of UV radiation,
but no disk contribution is seen.}

Is it plausible that the flat-disk assumption in eclipse mapping
studies might cause distances to be underestimated? Roughly
speaking, these estimates are based on the color and brightness of  
the pixels in a disk map, and assuming these to radiate as
blackbodies (BBs) or stellar atmospheres (SAs). If the disk is assumed
to be flat, the flux from each pixel then scales as $\cos{i} / d^2$
(modulo a correction for limb-darkening); this allows $d$~to be
estimated. However, most other accretion structures (e.g. a disk 
rim) would be {\em less} strongly foreshortened when averaged over 
the visible projected area of the structure. The flat-disk scaling 
will yield erroneous results in such cases. To see this, consider an
accretion flow element that is seen more face-on than the pixel 
representing it in a flat-disk eclipse map. The temperature of this
element/pixel is roughly fixed by its color. If a $\cos{i}$
foreshortening factor is incorrectly applied to this element/pixel by
the eclipse mapping algorithm, the distance inferred from it will be 
an underestimate. After all, the excessively foreshortened disk map
pixel must still produce the observed flux. This is only possible if
it is brought closer than the face-on accretion flow element it 
represents. Thus the flat disk assumption
might indeed cause distances to be underestimated systematically. When
this happens, the color-magnitude diagrams constructed from disk map
pixels will probably also show large scatter about the expected BB/SA
relations (since different accretion flow elements are likely to have
different foreshortening factors). This would appear to be the case in
DW~UMa (see Figure~5 in B{\'{\i}}r{\' o} 2000).

As a final note on the distance issue, we acknowledge that the
distance estimates based on Bailey's method and the WD-dominated  UV
spectrum are themselves only marginally consistent with each other (see
discussion in Araujo-Betancor et al. 2003). However, these estimates
are much more easily reconciled with each other than with the short
eclipse mapping distance. We have already shown explicitly that a WD
at 300~pc produces a much broader Lyman alpha feature than is
observed. But this distance would also require an unrealistically late 
spectral type for the secondary (to avoid over-predicting the
low-state, mid-eclipse K-band flux).

Turning to our detection of a UV counterpart to the phase 0.5
absorption events seen in optical spectra of SW~Sex stars, we note
that this could not necessarily have been expected: the formation
mechanism of these high-ionization (and  mostly resonance) lines might
be expected to differ significantly from that of the optical
low-ionization lines where the effect is normally seen. 
Strictly speaking, we cannot yet claim that the UV lines are affected
by genuine {\em absorption} near phase 0.5: none of the lines exhibit
dips below continuum level at this phase, so the data are also 
consistent with an intrinsic weakening of the lines. However, the
Balmer lines of SW~Sex stars generally also do not show absorption 
below continuum, although this effect has been seen in the He~{\sc i}
lines of several systems (e.g. Szkody \& Pich\'{e} 1990). On balance,
we are therefore inclined to ascribe the weakening of the UV lines to
the same absorbing medium that is affecting the optical lines. Given
the wide range of transitions affected, this probably implies
non-uniform physical conditions in the absorbing material.

\vspace*{0.1cm}
This work made use of Peter van Hoof's online atomic line list.
\footnote{\url{http://www.pa.uky.edu/\textasciitilde peter/atomic}}
Financial support for KSL, SAB, PS and DWH was provided by NASA
through grant numbers GO-9791 and GO-7362 from the Space Telescope
Science Institute (STScI), which is operated by AURA, Inc.,under NASA
contract NAS5-26555. RIH is supported by NASA through Hubble
Fellowship grant HF-01150.01-A awarded by STScI. BTG was supported by
a PPARC Advanced Fellowship.

\bibliographystyle{apj}
\bibliography{bibliography}

\begin{figure}
\figurenum{1}
\epsscale{0.68}
\plotone{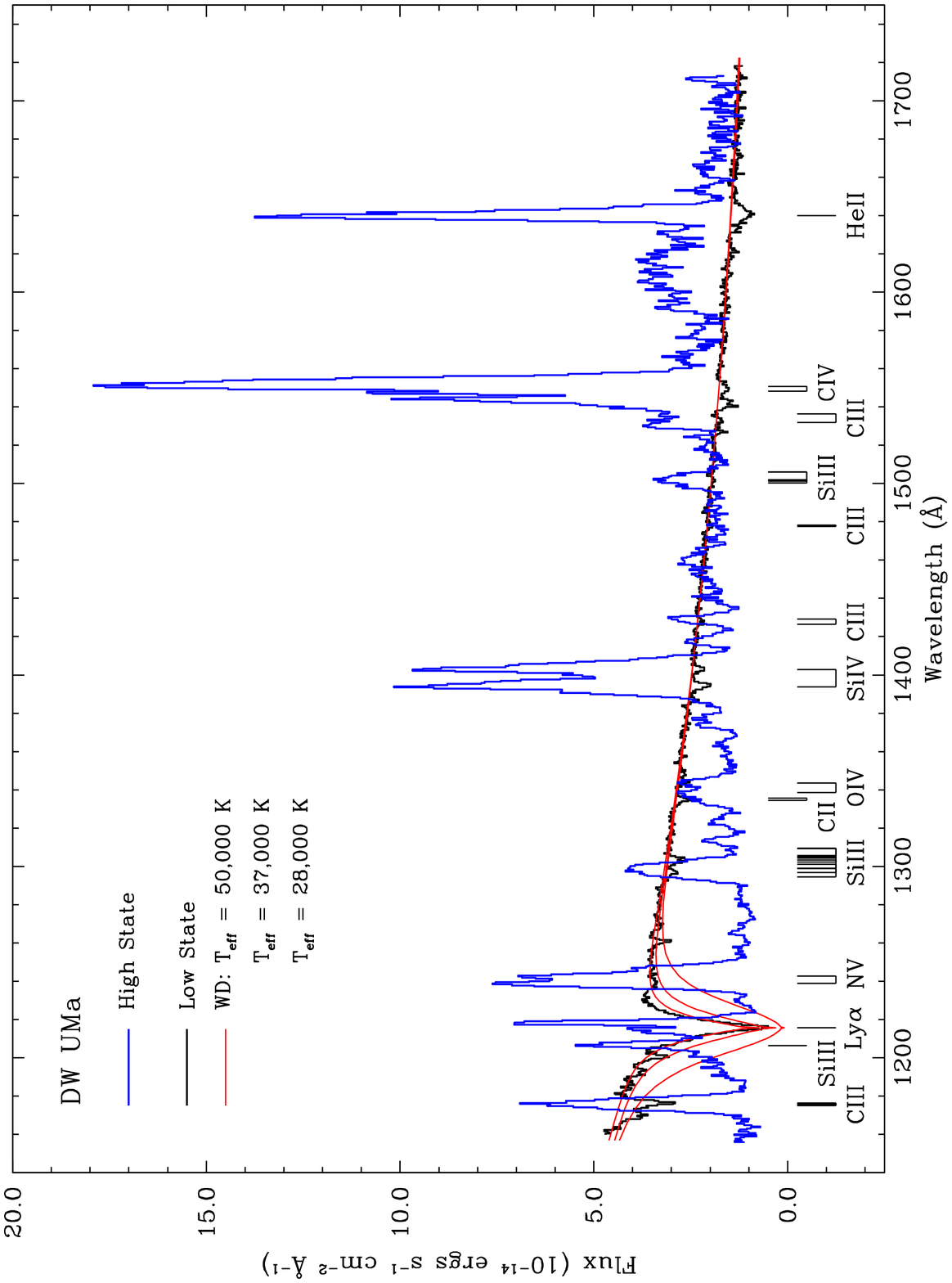}
\figcaption{DW~UMa's out-of-eclipse UV spectrum in the high (blue
line) and low state (black line). Suggested line identifications for
the strong features in the spectra are given. Also shown are three DA
WD models with different effective temperatures (red lines), all of
which have been normalized to match the flux of the low-state spectrum
at 1480~\AA. The effective temperatures of the models are indicated,
with cooler temperatures yielding spectra with broader Lyman~$\alpha$
features.}
\end{figure}

\begin{figure}
\figurenum{2}
\epsscale{0.68}
\plotone{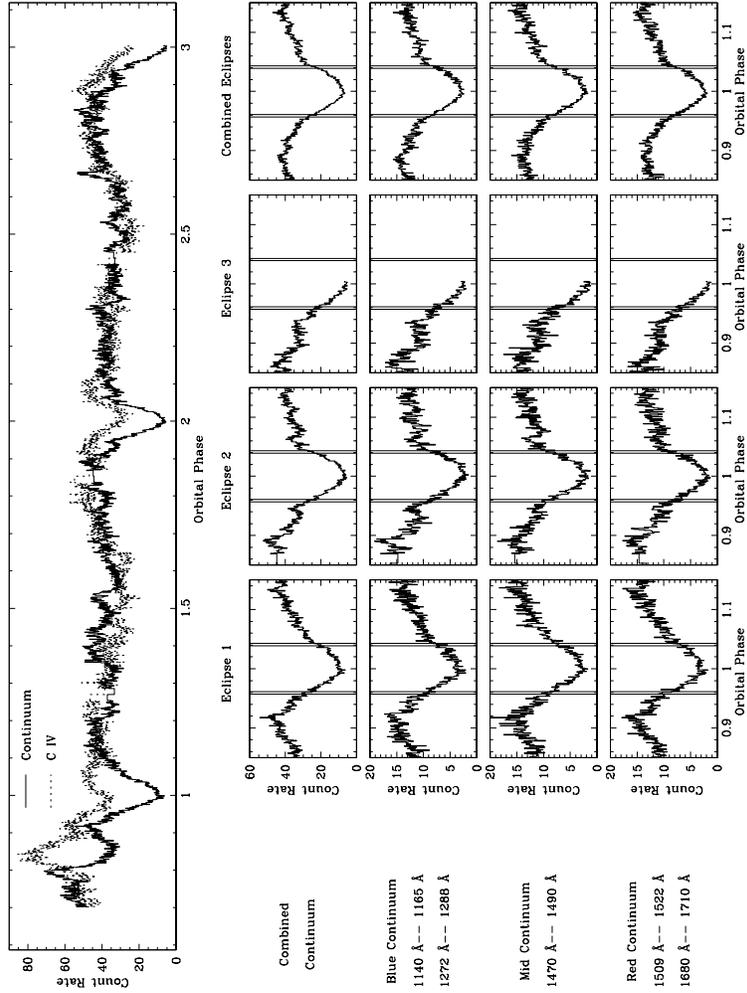}
\figcaption{The high-state, UV light curves. The top panel
shows continuum (continuous line) and C~{\sc iv} line flux (broken
line) light curves. The three bottom rows of panels show the
continuum eclipses in more detail. Each row corresponds to a
particular set of continuum windows (as indicated) and the eclipses are shown
individually (first three columns) and in combination (last
column). The vertical lines mark the WD contact phases measured by 
Araujo-Betancor et al. (2003). Note that the continuum lightcurve
in the top row is constructed from the summed count rates in all
continuum windows.}
\end{figure}

\begin{figure}
\figurenum{3}
\epsscale{0.68}
\plotone{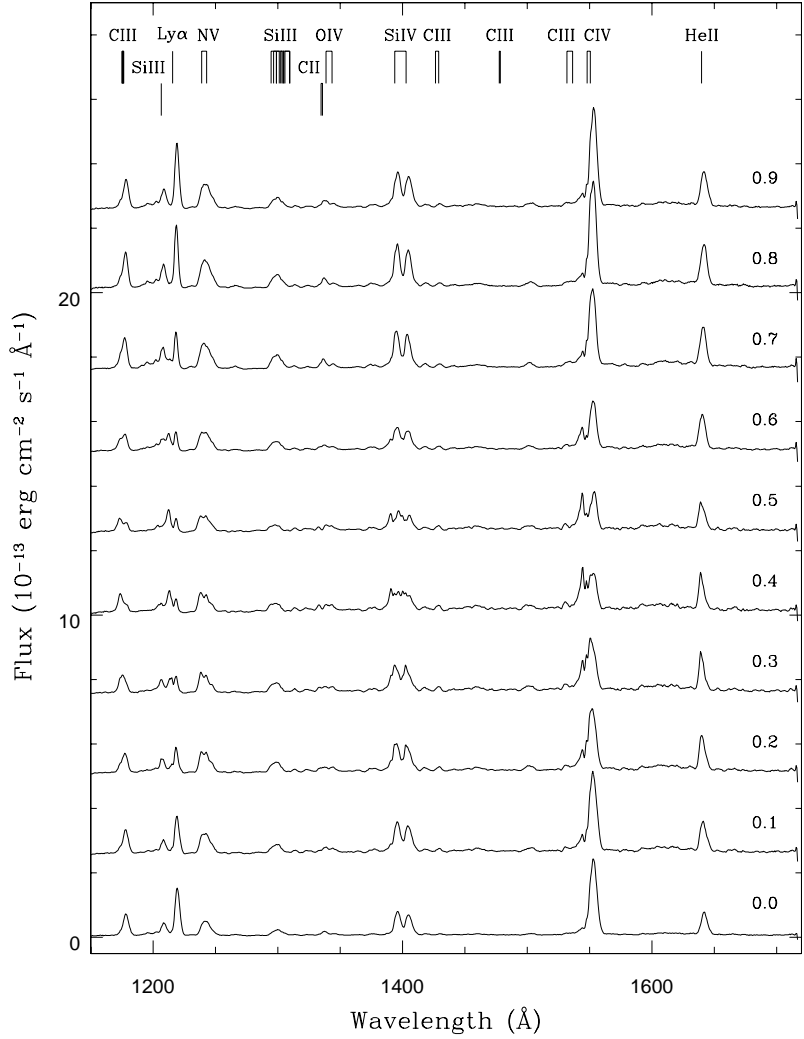}
\figcaption{Orbital-phase binned, high-state UV spectra. Only the bottom
(mid-eclipse) spectrum is on the correct flux scale, the others have
been offset by multiples of 2.5 in the units of the ordinate
axis. Suggested line identifications are marked, as are the orbital
phases corresponding to the mid-point of each bin.}
\end{figure}
 
\end{document}